\documentclass[11pt,a4paper]{article}

\PassOptionsToPackage{breaklinks}{hyperref}
\usepackage[hyperref]{acl2019}
\usepackage{times}
\usepackage{latexsym}

\usepackage{url}

\usepackage{booktabs}
\usepackage{graphicx}

\aclfinalcopy % Uncomment this line for the final submission
\title{Towards Multimodal Understanding of Passenger-Vehicle Interactions in Autonomous Vehicles: Intent/Slot Recognition Utilizing Audio-Visual Data}
\author{Eda Okur \qquad Shachi H Kumar \qquad Saurav Sahay \qquad Lama Nachman \\
  Intel Labs, Anticipatory Computing Lab, USA \\
  \texttt{\{eda.okur, shachi.h.kumar, saurav.sahay, lama.nachman\}@intel.com} \\}

\date{}

\begin{document}
\maketitle
%\begin{abstract}
%  This document contains the instructions for preparing %a camera-ready
%  manuscript for the proceedings of ACL 2019. The %document itself
%  conforms to its own specifications, and is therefore %an example of
%  what your manuscript should look like. These %instructions should be
%  used for both papers submitted for review and for %final versions of
%  accepted papers.  Authors are asked to conform to all %the directions
%  reported in this document.
%\end{abstract}

\section{Introduction}

Understanding passenger intents from spoken interactions and car's vision (both inside and outside the vehicle) are important building blocks towards developing contextual dialog systems for natural interactions in autonomous vehicles (AV). In this study, we continued exploring AMIE (Automated-vehicle Multimodal In-cabin Experience), the in-cabin agent responsible for handling certain multimodal passenger-vehicle interactions. When the passengers give instructions to AMIE, the agent should parse such commands properly considering available three modalities (language/text, audio, video) and trigger the appropriate functionality of the AV system. We had collected a multimodal in-cabin dataset with multi-turn dialogues between the passengers and AMIE using a Wizard-of-Oz scheme via realistic scavenger hunt game. %We focused on AMIE scenarios describing usages around setting/changing destinations/routes, updating driving behavior or speeds, finishing the trip and other use-cases to support various natural commands. 

In our previous explorations~\cite{AMIE-WiML-NIPS-2018, AMIE-CICLing-2019}, we experimented with various RNN-based models to detect utterance-level intents (set destination, change route, go faster, go slower, stop, park, pull over, drop off, open door, and others) along with intent keywords and relevant slots (location, position/direction, object, gesture/gaze, time-guidance, person) associated with the action to be performed in our AV scenarios. %After that, we continued our explorations with conducting speech-to-text experiments by comparing NLU models trained on human transcriptions versus noisy Automatic Speech Recognition (ASR) outputs and compared the results with single passenger rides versus the rides with multiple passengers. 

In this recent work, we propose to discuss the benefits of multimodal understanding of in-cabin utterances by incorporating verbal/language input (text and speech embeddings) together with the non-verbal/acoustic and visual input from inside and outside the vehicle (i.e., passenger gestures and gaze from in-cabin video stream, referred objects outside of the vehicle from the road view camera stream). Our experimental results outperformed text-only baselines and with multimodality, we achieved improved performances for utterance-level intent detection and slot filling.

\section {Methodology}

We explored leveraging multimodality for the NLU module in the SDS pipeline. As our AMIE in-cabin dataset\footnote{Details of AMIE data collection setup in~\cite{HRI-2018, AMIE-CICLing-2019}; in-cabin dataset statistics in~\ref{sec:appendix}.} has video and audio recordings, we investigated 3 modalities for the NLU: text, audio, and video. For text (language) modality, our previous work~\cite{AMIE-CICLing-2019} presents the details of our best-performing Hierarchical \& Joint Bi-LSTM models~\cite{bi-lstm-1997, hakkani-2016, zhang-2016, wen-2018} (H-Joint-2, see~\ref{sec:appendix}) and the results for utterance-level intent recognition and word-level slot filling via transcribed and recognized (ASR output) textual data, using word embeddings (GloVe~\cite{glove-2014}) as features. This study explores the following multimodal features:

%\subsubsection{Speech Embeddings}
\textbf{Speech Embeddings}: We incorporated pre-trained speech embeddings (Speech2Vec~\cite{speech2vec-Chung2018}) as features, trained on a corpus of~500 hours of speech from LibriSpeech. Speech2Vec\footnote{github.com/iamyuanchung/speech2vec-pretrained-vectors} is considered as a speech version of Word2Vec~\cite{word2vec-nips-2013} which is compared with Word2Vec vectors trained on the transcript of the same speech corpus. We experimented with concatenating word and speech embeddings by using pre-trained GloVe embeddings (6B tokens, 400K vocab, dim=100), Speech2Vec embeddings (37.6K vocab, dim=100), and its Word2Vec counterpart (37.6K vocab, dim=100). %Various feature concatenations and Precision/Recall/F1 (\%) results of NLU models can be found in Table 1 (in-cabin data, 1331 utterances).

\begin{table*}[t]
%\tiny
%\scriptsize
%\footnotesize
\small
\centering
\resizebox{0.78\textwidth}{!}{
\begin{tabular}{|c|l|c c c|c c c|}
\hline &  & \multicolumn{3}{c|}{\textbf{Intent Recognition}} & \multicolumn{3}{c|}{\textbf{Slot Filling}} \\
 % &  & \multicolumn{3}{c|}{\textbf{Utterance-Level}} & \multicolumn{3}{c|}{\textbf{Slot Filling \& Intent}} \\
 % &  & \multicolumn{3}{c|}{\textbf{Intent Recognition}} & \multicolumn{3}{c|}{\textbf{Keyword Extraction}} \\ 
 % &  & \multicolumn{3}{c|}{(Hierarchical \& Joint)} & \multicolumn{3}{c|}{(Joint Word-Level)} \\ 
 % \cmidrule(lr){3-5} \cmidrule(lr){6-8}
\textbf{Modalities} & \textbf{Features (Embeddings)} & Prec & Rec & F1 & Prec & Rec & F1 \\ \hline
Text & GloVe (400K) & 89.2 & 89.0 & 89.0 & 95.8 & 95.8 & 95.8 \\
Text & Word2Vec (37.6K) & 86.4 & 85.2 & 85.6 & 93.3 & 93.4 & 93.3 \\
Audio & Speech2Vec (37.6K) & 85.1 & 84.4 & 84.5 & 93.2 & 93.3 & 93.1 \\
Text \& Audio & Word2Vec + Speech2Vec & 88.4 & 88.1 & 88.1 & 94.2 & 94.3 & 94.2 \\
Text \& Audio & GloVe + Speech2Vec & 91.1 & 91.0 & 90.9 & 96.3 & 96.3 & 96.3 \\
Text \& Audio & GloVe + Word2Vec + Speech2Vec & 91.5 & 91.2 & 91.3 & 96.6 & 96.6 & 96.6 \\
\hline
\end{tabular}
}
\caption{\label{speech2vec} Speech Embeddings Experiments: Precision/Recall/F1-scores (\%) of NLU Models} %on in-cabin data (1331 utterances).}
\end{table*}

\begin{table*}[t]
%\tiny
%\scriptsize
%\footnotesize
\small
\centering
\resizebox{\textwidth}{!}{
\begin{tabular}{|c|l|c c c|}
\hline 
%&  & \multicolumn{3}{c|}{\textbf{Intent Recognition}} \\
\textbf{Modalities} & \textbf{Features} & Prec & Rec & F1 \\ \hline
Text & Embeddings (GloVe) & 89.19 & 89.04 & 89.02 \\
Text \& Audio & Embeddings (GloVe) + Audio (openSMILE/IS10) & 89.69 & 89.64 & 89.53 \\
Text \& Video & Embeddings (GloVe) + Video\_cabin (CNN/Inception-ResNet-v2) & 89.48 & 89.57 & 89.40 \\
Text \& Video & Embeddings (GloVe) + Video\_road (CNN/Inception-ResNet-v2) & 89.78 & 89.19 & 89.37 \\
Text \& Video & Embeddings (GloVe) + Video\_cabin+road (CNN/Inception-ResNet-v2) & 89.84 & 89.72 & 89.68 \\
\hline
Text \& Audio & Embeddings (GloVe+Word2Vec+Speech2Vec) & 91.50 & 91.24 & 91.29 \\
Text \& Audio & Embeddings (GloVe+Word2Vec+Speech2Vec) + Audio (openSMILE) & 91.83 & 91.62 & 91.68 \\
Text \& Audio \& Video & Embeddings (GloVe+Word2Vec+Speech2Vec) + Video\_cabin (CNN) & 91.73 & 91.47 & 91.50 \\
Text \& Audio \& Video & Embeddings (GloVe+Word2Vec+Speech2Vec) + Video\_cabin+road (CNN) & 91.73 & 91.54 & 91.55 \\
\hline
\end{tabular}
}
\caption{\label{audio-video} Multimodal (Audio \& Video) Features Exploration: Precision/Recall/F1-scores (\%) of Intent Recognition}
\end{table*}

\textbf{Audio Features}: Using openSMILE~\cite{openSMILE-2013}, 1582 audio features are extracted for each utterance using the segmented audio clips from in-cabin AMIE dataset. These are the INTERSPEECH 2010 Paralinguistic Challenge features (IS10) including PCM loudness, MFCC, log Mel Freq. Band, LSP, etc.~\cite{IS10-schuller2010interspeech}.

\textbf{Video Features}: Using the feature extraction process described in~\cite{cnn-feats-kordopatis2017}, we extracted intermediate CNN features\footnote{github.com/MKLab-ITI/intermediate-cnn-features} for each segmented video clip from AMIE dataset. For any given input video clip (segmented for each utterance), one frame per second is sampled and its visual descriptor is extracted from the activations of the intermediate convolution layers of a pre-trained CNN. We used the pre-trained Inception-ResNet-v2 model\footnote{github.com/tensorflow/models/tree/master/research/slim}~\cite{inception-SzegedyIV16} and generated 4096-dim features for each sample. We experimented with adding 2 sources of visual information: (i) cabin/passenger view from the BackDriver RGB camera recordings, (ii) road/outside view from the DashCam RGB video streams.

\section{Experimental Results}

For incorporating speech embeddings experiments, performance results of NLU models 
%using hierarchical joint learning)
on in-cabin data %(1331 utterances) 
with various feature concatenations can be found in Table~\ref{speech2vec}, using our previous hierarchical joint model (H-Joint-2). When used in isolation, Word2Vec and Speech2Vec achieves comparable performances, which cannot reach GloVe performance. This was expected as the pre-trained Speech2Vec vectors have lower vocabulary coverage than GloVe. Yet, we observed that concatenating GloVe + Speech2Vec, and further GloVe + Word2Vec + Speech2Vec yields better NLU results: F1-score increased from 0.89 to 0.91 for intent recognition, from 0.96 to 0.97 for slot filling. 

For multimodal (audio \& video) features exploration, performance results of the compared models with varying modality/feature concatenations can be found in Table~\ref{audio-video}. Since these audio/video features are extracted per utterance (on segmented audio \& video clips), we experimented with the utterance-level intent recognition task only, using hierarchical joint learning (H-Joint-2). We investigated the audio-visual feature additions on top of text-only and text+speech embedding models. Adding openSMILE/IS10 features from audio, as well as incorporating intermediate CNN/Inception-ResNet-v2 features from video brought slight improvements to our intent models, reaching 0.92 F1-score. These initial results using feature concatenations may need further explorations, especially for certain intent-types such as stop (audio intensity) or relevant slots such as passenger gestures/gaze (from cabin video) and outside objects (from road video).

\section{Conclusion}

In this study, we present our initial explorations towards multimodal understanding of passenger utterances in autonomous vehicles. We briefly show that our experimental results outperformed certain baselines and with multimodality, we achieved improved overall F1-scores of 0.92 for utterance-level intent detection and 0.97 for word-level slot filling. This ongoing research has a potential impact of exploring real-world challenges with human-vehicle-scene interactions for autonomous driving support with spoken utterances.%, which involves understanding real-world challenges with Spoken Interactions and Autonomous Driving.

\bibliography{acl2019}
\bibliographystyle{acl_natbib}

\appendix

\section{Appendices}
\label{sec:appendix}
%Appendices are material that can be read, and include lemmas, formulas, proofs, and tables that are not critical to the reading and understanding of the paper. Appendices should be \textbf{uploaded as supplementary material} when submitting the paper for review. Upon acceptance, the appendices come after the references, as shown here. Use \verb|\appendix| before any appendix section to switch the section numbering over to letters.

\textbf{AMIE In-cabin Dataset}: We obtained 1331 utterances having commands to AMIE agent from our in-cabin dataset. Annotation results for \textit{utterance-level intent} types, \textit{slots} and \textit{intent keywords} can be found in Table~\ref{D1} and Table~\ref{D2}.

%\addtolength{\tabcolsep}{+3pt} 
%\begin{table*}[t]
\begin{table}[h]
%\tiny
\scriptsize
%\footnotesize
%\small
  \centering
  %\resizebox{\columnwidth}{!}{
  \begin{tabular}{*3c}
    \toprule
    \textbf{AMIE Scenario} & \textbf{Intent Type} & \textbf{Utterance Count} \\
    \toprule
    Set/Change & SetDestination & 311 \\
    Destination/Route & SetRoute & 507 \\
    \midrule
     & Park & 151 \\
    Finishing the Trip & PullOver & 34 \\
     & Stop & 27 \\
    \midrule
    Set/Change & GoFaster & 73 \\
    Driving Behavior/Speed & GoSlower & 41 \\
    \midrule
    Others & OpenDoor & 136 \\
    (Door, Music, A/C, etc.) & Other & 51 \\
    \midrule
     & \textit{Total} & \textit{1331} \\
    \bottomrule
  \end{tabular}
  %}
  \caption{AMIE In-cabin Dataset Statistics: Intents}
  \label{D1}
\end{table}
%\addtolength{\tabcolsep}{-3pt} 

%\addtolength{\tabcolsep}{+4pt}
\begin{table}[h]
%\tiny
\scriptsize
%\footnotesize
%\small
  \centering
  %\resizebox{\columnwidth}{!}{
  \begin{tabular}{*2c}
    \toprule
    \textbf{Slot/Keyword Type} & \textbf{Word Count} \\
    \toprule
    Intent Keyword & 2007 \\
    %\hline
    \midrule
    Location & 1969 \\
    Position/Direction & 1131 \\
    Person & 404 \\
    Time Guidance & 246 \\
    Gesture/Gaze & 167 \\
    Object & 110 \\
    %None & 8519 \\
    %\hline
    \midrule
    None & 6512 \\
    \midrule
    \textit{Total} & \textit{12546} \\
    \bottomrule
  \end{tabular}
  %}
  \caption{AMIE In-cabin Dataset Statistics: Slots}
  \label{D2}
\end{table}
%\addtolength{\tabcolsep}{-4pt}

\textbf{Hierarchical \& Joint Model (H-Joint-2)}: 2-level hierarchical joint learning model that detects/extracts \textit{intent keywords \& slots} using seq2seq Bi-LSTMs first (Level-1), then only the words that are predicted as \textit{intent keywords \& valid slots} are fed into Joint-2 model (Level-2), which is another seq2seq Bi-LSTM network for \textit{utterance-level intent} detection (jointly trained with \textit{slots} \& \textit{intent keywords})~\cite{AMIE-CICLing-2019}.
%Level-1 (seq2seq Bi-LSTM for \textit{intent keywords} \& \textit{slots} extraction) + Level-2 
%(Joint-2 Seq2seq models with \textit{slots} \& \textit{intent keywords} \& \textit{utterance-level intents})%
%(Joint-2: seq2seq Bi-LSTM for \textit{utterance-level intent} detection (jointly trained with \textit{slots} \& \textit{intent keywords})) 

\end{document}